\documentclass{elsarticle}

\usepackage{graphicx}
\usepackage{amsmath}
\usepackage{amssymb}
\usepackage{xcolor}
\usepackage{hyperref}
\hypersetup{
    colorlinks=true,       
    linkcolor=red,          
    citecolor=blue,        
    filecolor=magenta,      
    urlcolor=cyan           
}
\usepackage{threeparttable}

\newcommand{\halley}{1P/Halley }
\newcommand{\halleyE}{1P/Halley}
\newcommand{\tuttle}{8P/Tuttle }
\newcommand{\tuttleE}{8P/Tuttle}
\newcommand{\borrelly}{19P/Borrelly }
\newcommand{\borrellyE}{19P/Borrelly}
\newcommand{\chury}{67P/Churyumov-Gerasimenko }
\newcommand{\churyE}{67P/Churyumov-Gerasimenko}
\newcommand{\hartley}{103P/Hartley }
\newcommand{\hartleyE}{103P/Hartley}

\journal{Icarus}

\biboptions{authoryear}

\begin{document}

\begin{frontmatter}

\title{Chaotic dynamics around cometary nuclei}

\author[besancon]{Jos\'e Lages}
\ead{jose.lages@utinam.cnrs.fr}
\author[besancon,pulkovo,moscow]{Ivan I. Shevchenko}
\ead{iis@gao.spb.ru}
\author[besancon]{Guillaume Rollin}
\ead{guillaume.rollin@utinam.cnrs.fr}

\address[besancon]{Institut UTINAM, Observatoire des Sciences de l'Univers THETA,
CNRS, Universit\'e de Bourgogne-Franche-Comt\'e, Besan\c{c}on 25030, France}
\address[pulkovo]{Pulkovo Observatory, RAS, 196140 Saint Petersburg, Russia}
\address[moscow]{Lebedev Physical Institute, RAS, 119991 Moscow, Russia}

\begin{abstract}
We apply a generalized Kepler map theory to describe the
qualitative chaotic dynamics around cometary nuclei, based on
accessible observational data for five comets whose nuclei are
well-documented to resemble dumb-bells. The sizes of chaotic zones
around the nuclei and the Lyapunov times of the motion inside
these zones are estimated. In the case of Comet \halleyE, the circumnuclear chaotic
zone seems to engulf an essential part of the Hill sphere, at
least for orbits of moderate to high eccentricity.
\end{abstract}


\end{frontmatter}


\section{Introduction}
\label{intro}

Asteroids and cometary nuclei quite often have bilobed (dumb-bell)
shapes; a spectacular example of such a shape is offered by the
recent radar imaging of the near-Earth asteroid 2014~JO25\footnote{\url{https://www.jpl.nasa.gov/news/news.php?feature=6817}}. The orbital dynamics around bodies with complex
gravity fields \citep{chauvineau93,SOH96,scheeres98,petit97,scheeres02,bartczak03,mysen06,olsen06,mysen07,feng17} and in particular around contact-binary solid
bodies \citep{marchis14,feng16}, was explored thoroughly in the last two decades
\citep[see a
brief review in][]{lages17}.

Here we use a generalized Kepler map technique \citep{lages17} to describe the
global dynamics around cometary nuclei known to be bilobed.
We recall that the Kepler map is a two-dimensional area-preserving
map describing the eccentric circumbinary motion of a massless
particle. The motion is described in terms of increments in
particle's energy and orbital period measured at its pericenter
and apocenter passages. The Kepler map was verified to be a
powerful tool to study resonant and chaotic orbital dynamics of
comets, in particular Comet Halley \citep{petrosky86,chirikov86,chirikov89,rollin15b}. One
should emphasize that here we apply this technique to describe the
orbital motion {\it around cometary nuclei}, in particular around
the nucleus of Comet Halley. This outlines the general character
of the technique. Also the Kepler map has been applied to very different domains, from strong microwave ionization of
excited hydrogen atoms \citep{casati88} and autoionization of molecular Rydberg states \citep{benvenuto94} to capture of dark matter by the solar system \citep{lages13} and by gravitating binaries in general \citep{rollin15a}, also to description of transfer trajectories of spacecrafts \citep{ross07}.

The employed Kepler map formalism allows one to straightforwardly estimate the characteristics of the chaotic zones around the cometary nuclei using simple analytical formulas, avoiding numerical integrations. What is more, in contrast to numerical integrations, it provides a direct physical insight in the dynamical problem of circumbinary motion: one is able to directly see which resonances overlap or interact; which Lyapunov timescales can be expected; how the chaotic diffusion in the energy variable may proceed; etc.

In \cite{lages17}, the Kepler map has been generalized to
describe the motion of a massless particle in the gravitational
field of a rotating irregularly shaped body modelled by a non-symmetric
dumb-bell. This generalization was achieved by introduction of an additional
parameter, $\omega$, responsible for the arbitrary rate of
rotation of the ``central binary''. Analytical expressions for the
coefficients of the ``kick function'', representing the energy
increment for the test particle per orbital revolution, were
derived.

In this article, we use the new technique introduced in
\cite{lages17} to describe the qualitative chaotic dynamics
around bilobed cometary nuclei, based on the data for five comets
whose nuclei are well documented to be dumb-bells. Note that the
majority of cometary nuclei whose shapes are well documented are
in fact dumb-bells. The sizes of chaotic zones around the nuclei
and the Lyapunov times of the motion inside these zones are
estimated. The nucleus of Comet Halley is addressed in particular;
we show that, due to its relatively slow rotation, a huge zone of
chaos is generated around this object.

\section{Bilobed cometary nuclei}
\label{dbac}

\begin{table}[t]
\begin{center}
\caption{\label{Table_data}
Bilobed cometary nuclei: basic observational data}
\begin{threeparttable}
\begin{tabular}{lrrrrl}
\noalign{\medskip} \hline \noalign{\medskip}
Comet & $d$, km & $M$, kg & $m_1/m_2$ & $P$, h & Refs. \\
\noalign{\medskip} \hline\hline \noalign{\medskip}
67P/C--G       & 2.62 & $9.982 \times 10^{12}$ & 3.4 & 12.404 & \cite{jorda16} \\
\noalign{\medskip} \hline \noalign{\medskip}
\halley      & 7.7  & $2.2 \times 10^{14}$ & 2.6 & 176.4 & \cite{mereny90} \\
&&&&&\cite{stooke91} \\
&&&&&\cite{schleicher15} \\\noalign{\medskip} \hline \noalign{\medskip}
\tuttle      & 5.0  & $4 \times 10^{14}$\tnote{ a}  & 2.14 & 11.4 & \cite{harmon10} \\
\noalign{\medskip} \hline \noalign{\medskip}
\borrelly   & 4.0  & $2 \times 10^{13} $ & 3.5 & 25 & \cite{lamy98} \\
&&&&&\cite{soderblom04} \\
&&&&&\cite{oberst04} \\
&&&&&\cite{buratti04} \\
&&&&&\cite{britt04} \\
\noalign{\medskip} \hline \noalign{\medskip}
\hartley   & 1.2 & $2.2 \times 10^{11}$ & 3.3 & 18.2 & \cite{harmon11} \\
&&&&&\cite{thomas13} \\
&&&&&\cite{belton13} \\
\noalign{\smallskip} \hline
\end{tabular}
\begin{tablenotes}
\item[a]{\small In the absence of observational data, the mean mass density of \tuttle has been assumed to be
$0.5$~g~cm$^{-3}$, equal to the mean density of 67P/C-G.}
\end{tablenotes}
\end{threeparttable}
\end{center}
\end{table}

Many minor bodies in the Solar system are ``potato''-shaped,
roughly resembling ellipsoids, but it is not infrequent that
asteroids and cometary nuclei resemble dumb-bells, i.e., they are
more like dumb-bells than ellipsoids. Therefore, one can describe
the body as a dumb-bell straightforwardly. Remarkably, this is the
case for 5 comets out of 8 that have the shapes of their nuclei
well-documented \citep{jorda16}.

A well-known example is the nucleus of Comet
67P/Churyumov--Gerasi\-menko, the target of the Rosetta mission
\citep{jorda16}. Tantalizingly, Comet \halleyE, famous for its
well-studied chaotic orbital dynamics and which was the first
object studied within the Kepler map formalism \citep{chirikov86,chirikov89}, also
has a dumb-bell nucleus \citep{mereny90,stooke91}. Other comets with
well-documented dumb-bell nuclei are \tuttle \citep{harmon10},
\borrelly \citep{oberst04}, and \hartley \citep{thomas13}. Such
peculiar appearance is explained by non-uniform erosion of the
surface of a cometary nucleus when it passes close to the Sun
\citep{jewitt03}, or, alternatively, by ``soft collisions'' of single bodies
\citep{rickman15}. Table~\ref{Table_data} summarizes the basic observational data about the above cited cometary nuclei:
$d$ is the separation of the centers of
mass of the lobes; $m_2 / m_1$ is the ratio of masses of the
lobes ($m_2 < m_1$); $M$ is the total mass of the nucleus; $P$ is
the observed rotation period of the nucleus.

Let us take the example of the nucleus of Comet 67P/Churyumov--Gerasi\-menko which from the observational data
presented in \cite{jorda16} can be described as an
aggregate of two merged bodies with the ratio of masses $m_1 / m_2
\simeq 3.4$; and their centers of mass are separated by $d
\simeq 2.62$~km. The total mass and the rotation period of the
nucleus are, respectively, $9.982 \times 10^{12}$~Kg \citep{patzold16} and
$12.404$~h \citep{jorda16}.

The Keplerian rate of rotation of a binary with masses $m_1$, $m_2$, and size $d$ is straightforwardly calculated using the third Kepler's law (see, e.g., \cite{murray99}, eq. (2.22)).
Then,
it is straightforward to calculate that the rotation rate $\omega$ of Comet 67P/Churyumov--Gerasi\-menko satisfies
the relation $\omega \simeq 0.73\omega_0$. Analogously, we have $\omega/\omega_0\simeq0.055$, $0.33$, $0.48$, and $1.04$ for \halleyE, \tuttleE, \borrellyE, and \hartleyE. Therefore, for the studied cometary nuclei, the rotation rate $\omega$ ranges from $0.055\omega_0$ to $1.04\omega_0$. However note that $\omega > \omega_0$ are not generally probable, as a rubble-pile object would disintegrate at such high rates of rotation.

\section{The dumb-bell map}
\label{ddbm}

For the clarity of the subsequent presentation, let us review the
Kepler map techniques.

In \cite{petrosky86} and \cite{chirikov86,chirikov89}, the classical Kepler map was introduced
as a tool for description of the chaotic motion of comets in
eccentric orbits. The model consists in the assumption that the
main perturbing effect of a planet is concentrated when the comet
is close to the perihelion of its orbit. This effect is defined by
the phase of encounter with the planet. The Kepler map has a
single parameter. Its analytical formula was first derived in the
restricted planar three-body problem framework in \cite{petrosky86} and \cite{petrosky88}. In
\cite{shevchenko11}, it was demonstrated that the Kepler map,
including analytical formulae for its parameter, can be derived by
quite elementary methods, based on the Jacobi integral formalism.
Following a procedure analogous to that presented in
\cite{shevchenko11}, the dumb-bell map can be straightforwardly
derived by allowing for an arbitrary rate of rotation of the
``central binary'' \citep{lages17}.

Consider the motion of a passively gravitating particle
in the planar restricted three-body problem
``$m_1$--$m_2$--particle'', where the two masses $m_1 \gg m_2$ are
connected by a massless rigid rod, thus forming an asymmetric
dumb-bell. Note that further on we consider solely the case of
prograde (with respect to the dumb-bell rotation) orbits of the
particle; analysis of the retrograde case is analogous.

We choose an inertial Cartesian coordinate system with the origin
at the dumb-bell's center of mass. Unless otherwise stated we express physical quantities is the following units:
the dumb-bell size $d$, i.e., the distance between centers of mass of the $m_1$ and $m_2$ lobes, is set to equal to unity, $d=1$;
we set the product $\mathcal{G}(m_1+m_2)=1$; consequently the angular frequency of the Keplerian orbital motion of the two lobes (i.e., the motion if the two masses $m_1$ and $m_2$ were unbound) is $\omega_0=\sqrt{\mathcal{G}(m_1+m_2)/d^3}=1$.
The motion of the particle with coordinates $(x, y)$ is
described by the differential equations
\begin{equation}
\ddot x = \nu \frac{x_1 - x}{r^3_{13}} + \mu \frac{x_2 - x}{r^3_{23}} ,
\qquad
\ddot y = \nu \frac{y_1 - y}{r^3_{13}} + \mu \frac{y_2 -
y}{r^3_{23}} ,
\label{diffeqs}
\end{equation}
with
\begin{equation}
\begin{array}{llrlllr}
x_1 &=& - \mu \cos[ \omega(t - t_0) ] ,&\qquad& x_2 &=& \nu \cos[ \omega(t - t_0) ] , \nonumber \\
y_1 &=& - \mu \sin[ \omega(t - t_0) ] ,&\qquad& y_2 &=& \nu \sin[ \omega(t - t_0) ] ,
\end{array}
\label{xyS}
\end{equation}
where $(x_1, y_1)$
and $(x_2, y_2)$ are the coordinates of $m_1$ and $m_2$,
respectively; $r_{i3} = \sqrt{(x_i - x)^2 + (y_i - y)^2}$ is the distance between the particle and the $m_i$ mass with $i=1,2$; 
$\mu = m_2/(m_1 + m_2)$ is the reduced mass of
$m_2$, $\nu = 1 - \mu$ is the reduced mass of $m_1$; and $t_0$ is an arbitrary time fixing the phase of the dumb-bell at $t=0$.
The quantity $\omega$ is an additional parameter, with respect to the usual equations of motion in the planar restricted three-body problem; this parameter is responsible for the arbitrary rotation frequency of the dumb-bell.

From (\ref{diffeqs}) and following either the methodology exposed in \cite{shevchenko11} or the potential gravity based methodology exposed in \cite{roy03}, it is possible to obtain  \citep{lages17}, for any rate of revolution of the central binary $\omega$ and any reduced mass $\mu$, the energy gain of the particle after a passage at the pericenter. The expression of the particle's energy kick function reads \citep{lages17}
\begin{equation}
\Delta E\left(\mu,q,\omega,\phi\right)
\simeq
W_1\left(\mu,q,\omega\right)\sin\left(\phi\right)+W_2\left(\mu,q,\omega\right)\sin\left(2\phi\right),
\label{deltaEall}
\end{equation}
where $\phi$ is the phase of the dumb-bell when particle is at pericenter, $q$ is the pericenter distance, and where the first harmonic coefficient reads
\begin{equation}\label{W1}
W_1\left(\mu,q,\omega\right)\simeq \mu\nu(\nu-\mu)2^{1/4}
\pi^{1/2} \omega^{5/2}
q^{-1/4} \exp \left( - \frac{2^{3/2}}{3} \omega q^{3/2} \right),
\end{equation}
and the second harmonic coefficient reads
\begin{equation}\label{W2}
W_2\left(\mu,q,\omega\right)\simeq-\mu\nu2^{15/4} \pi^{1/2} \omega^{5/2}q^{3/4}\exp \left( -
 \frac{2^{5/2}}{3} \omega q^{3/2} \right).
\end{equation}

The quasi-constancy of $q$ is an important issue considered, in particular, by \cite{shevchenko15} in the framework of the Jacobi constant formalism. At $a\gg d$, where $a$ is the orbiting particle's semimajor axis (measured in the units of the semimajor axis of the central binary), the pericentric distance $q$ is practically constant.

In the framework of the restricted three-body problem, if one
writes down the expression for the tertiary's energy increment
$\Delta E$ together with the expression for the increment of
perturber's phase angle $\phi$ between two consecutive passages of the tertiary at the pericenter, one obtains the usual Kepler map \citep{chirikov86,chirikov89,petrosky86}. In
the case of arbitrary rate of rotation of the central binary, the
map takes the form \citep{lages17}
\begin{equation}
\begin{array}{lll}
E_{i+1} &=& E_{i} + \Delta E\left(\mu,q,\omega,\phi_i\right), \\
\phi_{i+1} &=& \phi_{i} + 2 \pi \omega \vert 2 E_{i+1} \vert^{-3/2},
\label{kmp_gm2}
\end{array}
\end{equation}
where $E_i$ is the particle's energy at the $i$th passage at apocenter, and $\phi_i$ is the dumb-bell phase at the $i$th passage of the particle at pericenter.

Apart from the analysis of the global dynamical behavior, the Kepler map can be used to reproduce individual trajectories of a modeled dynamical system, but only to a certain extent. Generally, a comparative analysis of any map's performance versus a corresponding direct numerical integration for an individual trajectory does not make sense on the time intervals much greater than the trajectory's Lyapunov time $T_\mathrm{L}$, due to the essential sensitive dependence of the chaotic trajectories on the initial conditions. Therefore, in the case of the Kepler map, such a comparison for any chaotic trajectory does not make sense at all, because, according to \cite{shevchenko07}, the Kepler map's $T_\mathrm{L} \sim 1$ in the units of the map iterations; thus, any chaotic trajectories computed by different methods, though with the same initial conditions, would substantially diverge already on the timescale of several map iterations. (Of course, this divergence does not influence any comparison of the global dynamical behaviors, which should be the same statistically.) On the other hand, it is worthwhile to check the performance of the Kepler map, versus a direct numerical integration, taking as initial conditions those for an individual {\it regular} trajectory. Our computations show that a good accordance is observed indeed in this case: in particular, the deviations in the orbital energy do not exceed $0.1\%$ on the time intervals as high as $10^3$ periods of the binary; what is more, they do not increase but just oscillate. One should outline that the problem of numerical precision of the Kepler map in reproducing of individual trajectories has not yet been studied at all (though many authors used it to study statistics of chaotic trajectories), to our knowledge; this issue deserves a thorough separate study.

\section{Borders of chaos domain}
\label{chaosborder}

Let us estimate the size of the chaotic zone generated by the
rotating gravitating dumb-bell-shaped body, in application to
cometary nuclei. The analytical method for this estimation \citep{shevchenko15} is
based on the Kepler map approach developed for gravitating
non-bound binaries, such as, e.g., binary stars.
In \cite{shevchenko15}, analytical expressions for the energy kick functions $\Delta E$ and then estimations for chaos borders have been obtained in the cases of:
(a) highly asym\-me\-tric binaries, $\mu\ll1$, giving
$\Delta E(\mu,q,\phi)\simeq W(\mu,q)\sin\left(\phi\right)$ with $W(\mu,q)=W_1(\mu\ll1,q,\omega_0=1)$
and
(b) equal mass binaries, $\mu=1/2$, giving
$\Delta E(q,\phi)\simeq W(q)\sin\left(2\phi\right)$ with $W(q)=W_2(\mu=1/2,q,\omega_0=1)$.

For the five cometary nuclei, presented in Section~\ref{dbac}, and considered as contact solid binaries, the reduced mass $\mu$ is moderate and does not vary much from a nucleus to another, since $\mu=1/(1+m_1/m_2)$ ranges from $\approx0.22$ (for 19P/Borrely) to $\approx0.32$ (for \tuttle) (see Table~\ref{Table_data}). The cometary nuclei rotation rate $\omega$ can be very different from the Keplerian angular frequency $\omega_0$, since $\omega$ ranges from $\omega\ll\omega_0$ to $\omega\simeq\omega_0$ (see Table~\ref{Table_res}). Combining (\ref{W1}) and (\ref{W2}), one has
$W_1(\mu,q,\omega)/W_2(\mu,q,\omega)\simeq-\left(\nu-\mu\right)2^{-7/2}q^{-1}\exp \left(
\frac{2^{3/2}}{3} \omega q^{3/2} \right)$. For a pericenter distance greater than the dumb-bell size, $q>d$, we observe that the amplitude $W_2$ dominates over the amplitude $W_1$ for the angular frequencies of rotation such that $\omega/\omega_0\ll \left(d/q\right)^{3/2}$. In this regime of slow rotation, the border of chaos has been analytically estimated for rotating contact solid binaries \citep{lages17} as
\begin{equation}
E_\mathrm{cr}=-\Delta E_\mathrm{cr}\simeq-A(\mu\nu)^{2/5}\omega^{7/5}q^{3/10}\exp\left(-B\omega q^{3/2}\right),
\label{DE_dbm_W2}
\end{equation}
where $A = 2^{13/10} 3^{2/5} \pi^{3/5} K^{-2/5}$ and $B= 2^{7/2} / 15$. The width of the chaotic component around the separatrix (situated at $E=0$) is $\Delta E_\mathrm{cr}$; the particles in the chaotic component can not dynamically diffuse below $E<E_\mathrm{cr}$.
Conversely, for $\omega/\omega_0\gg \left(d/q\right)^{3/2}$, $W_1$ dominates over $W_2$. For the sake of completeness, we give here an estimate of the border of chaos in this regime of fast rotation. Consequently, dropping the second harmonic term in (\ref{deltaEall}) and using the
substitution $E = W_1 y$, $\phi = x$, the map~(\ref{kmp_gm2}) is reducible to
\begin{equation}
\begin{array}{lll}
y_{i+1} &=& y_i + \sin x_i,\\
x_{i+1} &=& x_i + \lambda \vert y_{i+1} \vert^{-3/2} ,
\end{array}
\label{kmp_gm1}
\end{equation}
where
\begin{equation}
\lambda = 2^{-1/2} \pi \omega W_1^{-3/2} . \label{kmp_la}
\end{equation}
As the dumb-bell map (\ref{kmp_gm1}) can be, locally in $(x,y)$ phase space, approximated by the standard map \citep{chirikov79,shevchenko14}, one finds for the location of the chaos border
\begin{equation}
y_\mathrm{cr} = \left( \frac{3 \lambda}{2 K_G} \right)^{2/5} ,
\label{ycr}
\end{equation}
where $K_G=0.971635406\dots$ \citep{shevchenko07}.
Using Eqs.~(\ref{W1}) and (\ref{kmp_la}) for $W_1$ and $\lambda$,
one obtains the half-width of the chaotic layer around the separatrix $(y=0)$
\begin{equation}
\Delta E_\mathrm{cr} = \left| E_\mathrm{cr} \right| = \left| W
y_\mathrm{cr}  \right| \simeq A [\mu\nu(\nu-\mu)]^{2/5}
\omega^{7/5} q^{-1/10} \exp \left(- B \omega q^{3/2}\right) ,
\label{DE_dbm}
\end{equation}
where $A = 2^{-1/2} 3^{2/5} \pi^{3/5} K_G^{-2/5}$ and $B= 2^{5/2} / 15$.
The particle's critical eccentricity $e_\mathrm{cr}$,
following from the relation $\Delta E_\mathrm{cr} = -
E_\mathrm{cr} = 1/2 a_\mathrm{cr} = (1 - e_\mathrm{cr})/2 q$,
is
\begin{equation}
e_\mathrm{cr} = 1 - 2 q \Delta E_\mathrm{cr} , \label{Km_ecr}
\end{equation}
where $\Delta E_\mathrm{cr}$ is given by
Eq.~(\ref{DE_dbm}). The orbits with $e \gtrsim
e_\mathrm{cr}(\omega, q)$ are chaotic.
Further on, the critical curve (\ref{Km_ecr}) will be superimposed on constructed stability diagrams. 



\begin{figure}
\begin{center}
\includegraphics[width=0.45\textwidth]{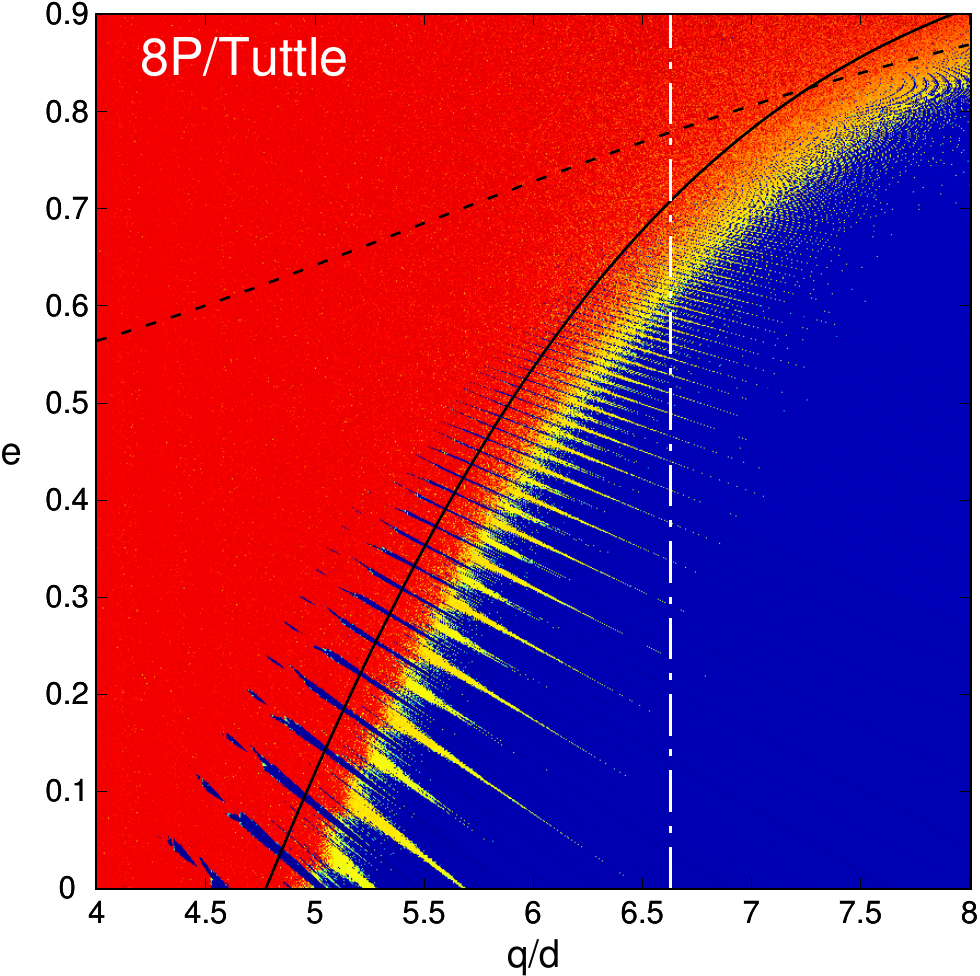}
\includegraphics[width=0.45\textwidth]{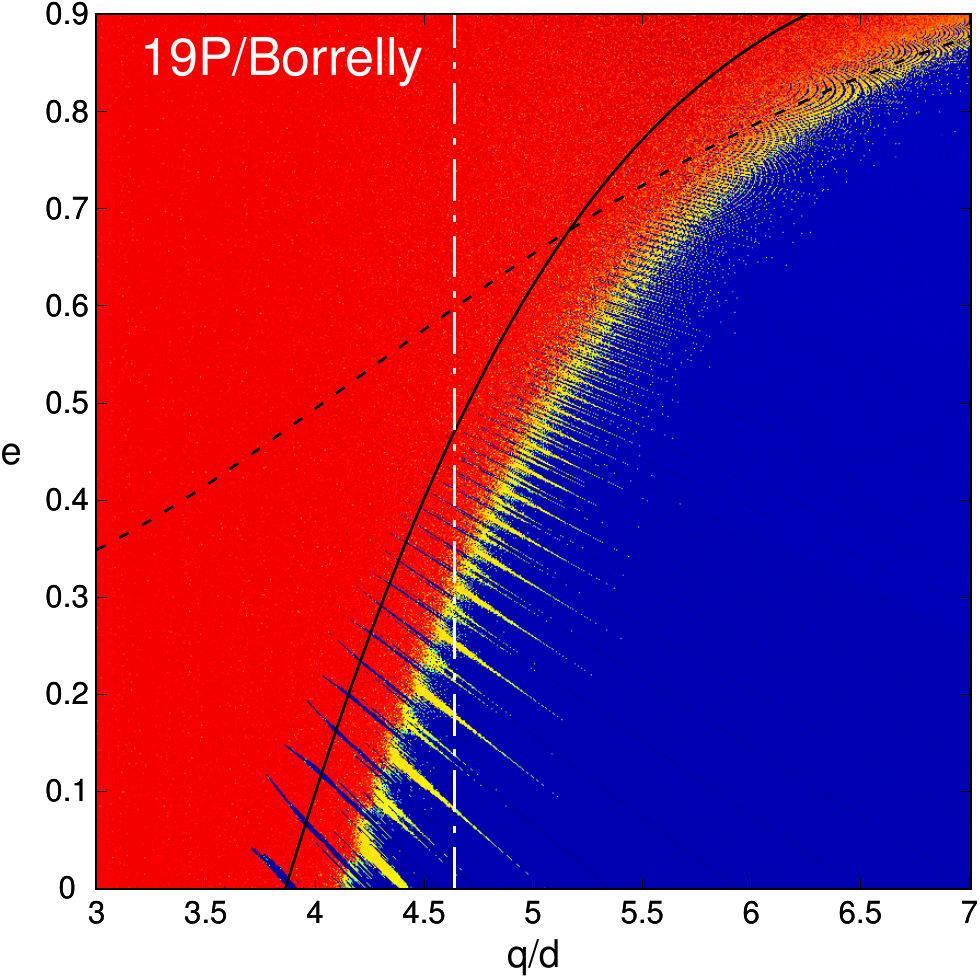}
\includegraphics[width=0.45\textwidth]{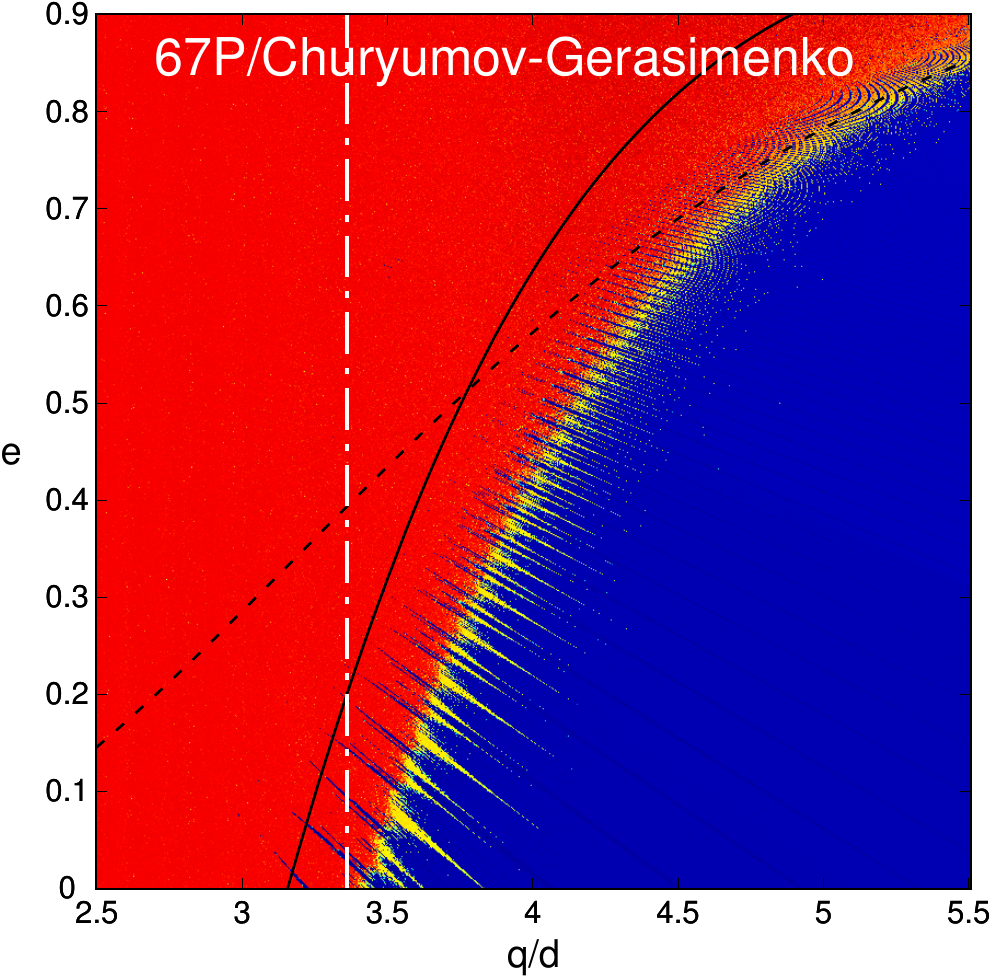}
\includegraphics[width=0.45\textwidth]{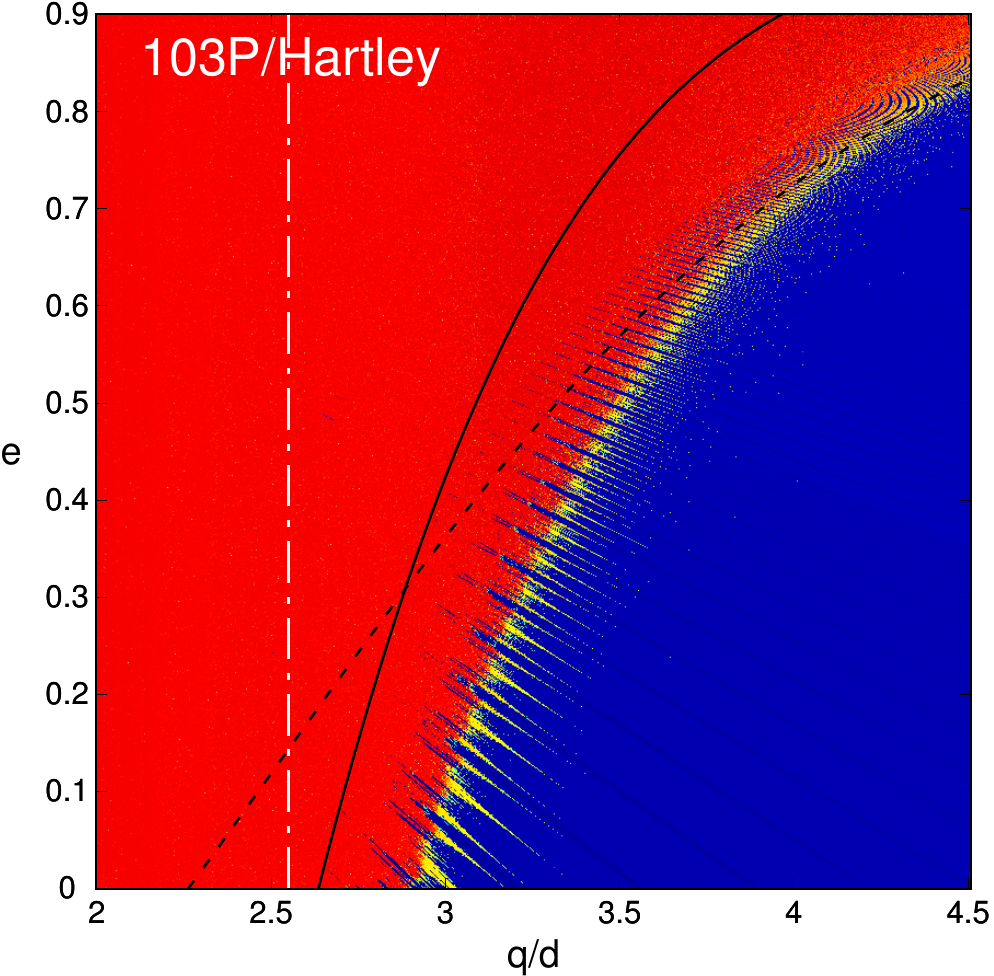}
\vskip 0.3cm
\includegraphics[width=0.45\textwidth]{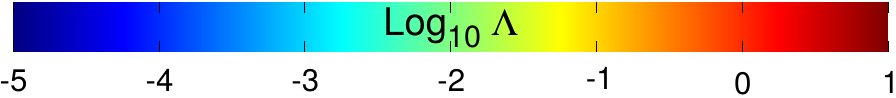}
\end{center}
\caption{Stability diagrams of the orbital motion around
four cometary nuclei ordered by increasing $\omega$:
\tuttle (top left, $\omega\simeq0.33\omega_0$),
\borrelly (top right, $\omega\simeq0.48\omega_0$),
\chury (bottom left, $\omega\simeq0.73\omega_0$),
\hartley (bottom right, $\omega\simeq1.04\omega_0$).
The chaotic domain is shown by the reddish area. Chaos is
determined by computing the maximum Lyapunov exponent $\Lambda$
for a trajectory with initial orbital elements $(q,e)$. Number of
iterations is $10^6$. The critical curve (\ref{Km_ecr}) is shown
taking into account only the first harmonic term in $\Delta E$ (\ref{deltaEall}) with amplitude
$W_1$ (\ref{W1}) (dashed black line) and taking into account
only the second harmonic in $\Delta E$ (\ref{deltaEall}) with amplitude $W_2$ (\ref{W2})
(solid black line). The vertical white dash-dotted line marks
the pericenter $q_0$ at which
$W_2=W_1$. For $q<q_0$
($q>q_0$), $W_2>W_1$ ($W_2<W_1$). The ratio $W_2/W_1$ ranges
from $10.2$ at $q=4d$ to $0.211$ at $q=8d$ for \tuttleE,
from $5.72$ at $q=3d$ to $0.0305$ at $q=7d$ for \borrellyE,
from $3.4$ at $q=2.5d$ to $0.0155$ at $q=5.5d$ for \churyE,
from $2.64$ at $q=2d$ to $0.00801$ at $q=4.5d$ for \hartleyE.
} \label{fig1}
\end{figure}

As mentioned already in the comment to Eqs. (\ref{kmp_gm1})-(\ref{kmp_la}), the Kepler map can be linearized in the energy variable so that to approximate the motion locally by the standard map.  This procedure provides powerful analytical means to estimate the local properties of the chaotic motion: local diffusion rates, local Lyapunov timescales, proximity to resonances, etc.

\section{Dynamics around nuclei of Comets}

\subsection{Stability diagrams}
\label{stabdiag}

Let us construct stability diagrams, using the dumb-bell map (\ref{kmp_gm2}), by computing the Lyapunov exponents on a fine grid of
initial conditions $(e,q)$. The Lyapunov exponents are calculated by means of iterating concurrently the dumb-bell map and its tangent
map \citep[see the general method described in][]{chirikov79}.
Fig.~\ref{fig1} presents the stability diagrams for particles orbiting cometary nuclei of \tuttleE, \borrellyE, \chury and 103P/Har\-tley whose physical parameters are given in Section~\ref{dbac}. For any given initial conditions $(e,q)$, the motion is regarded as chaotic if the maximum Lyapunov exponent $\Lambda$ is non-zero.
From Fig.~\ref{fig1} we see that a common peculiarity is that the border delimiting the chaotic domain (reddish area) and the regular domain (bluish area) is ragged. Here the most prominent \textit{teeth of instability} correspond to integer $p$:1 and half-integer $p+\frac12$:1 resonances. The prominence of the both integer and half-integer \textit{teeth} at the ragged border, well visible, e.g., in the \tuttle case, is explained by the fact that the two lobes of the considered cometary nuclei are comparable, i.e., $m_1\sim m_2$.


On the stability diagrams (Fig.~\ref{fig1}), we superimpose the critical curves $e_\mathrm{cr}(q)$ (\ref{Km_ecr}) computed for the dynamical regimes with $\omega/\omega_0\gg\left(d/q\right)^{3/2}$, using (\ref{DE_dbm}) as the expression for $\Delta E_\mathrm{cr}$ in (\ref{Km_ecr}), and $\omega/\omega_0\ll\left(d/q\right)^{3/2}$, using (\ref{DE_dbm_W2}) as the expression for $\Delta E_\mathrm{cr}$ in (\ref{Km_ecr}). From Fig.~\ref{fig1}, we clearly see that the critical curve for $\omega/\omega_0\gg\left(d/q\right)^{3/2}$ (dashed line) approximately describes the smoothed border of the chaotic domain at large pericenter distances $q$ and high eccentricities $e$. Conversely,   the critical curve for $\omega/\omega_0\ll\left(d/q\right)^{3/2}$ (solid line) approximately describes the smoothed border of the chaotic domain at lower pericenter distances and small eccentricities. These agreements testify the adequacy of the analytical approximation for chaos border (Section~\ref{chaosborder}). However we note that this approximation becomes less accurate at $\omega\gtrsim\omega_0$ as e.g., illustrated by the \hartley case.

\begin{figure}
\begin{center}
\includegraphics[width=0.8\textwidth]{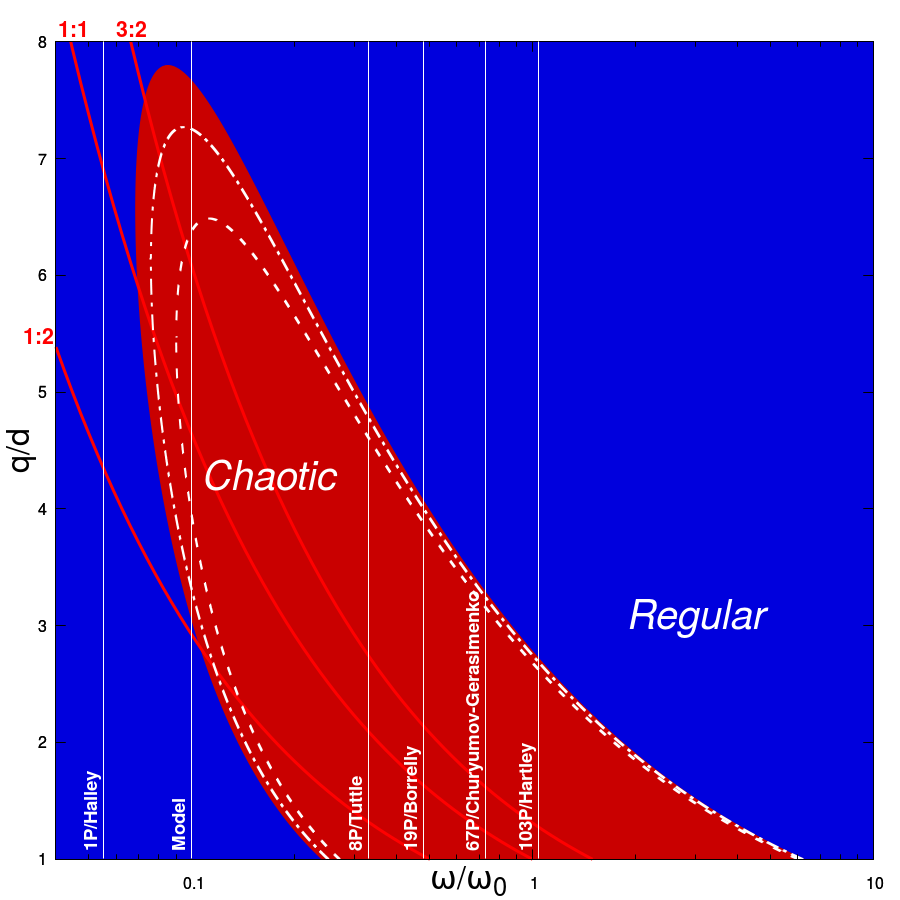}
\vskip 0.5cm \hskip 0.cm
\includegraphics[width=0.29\textwidth]{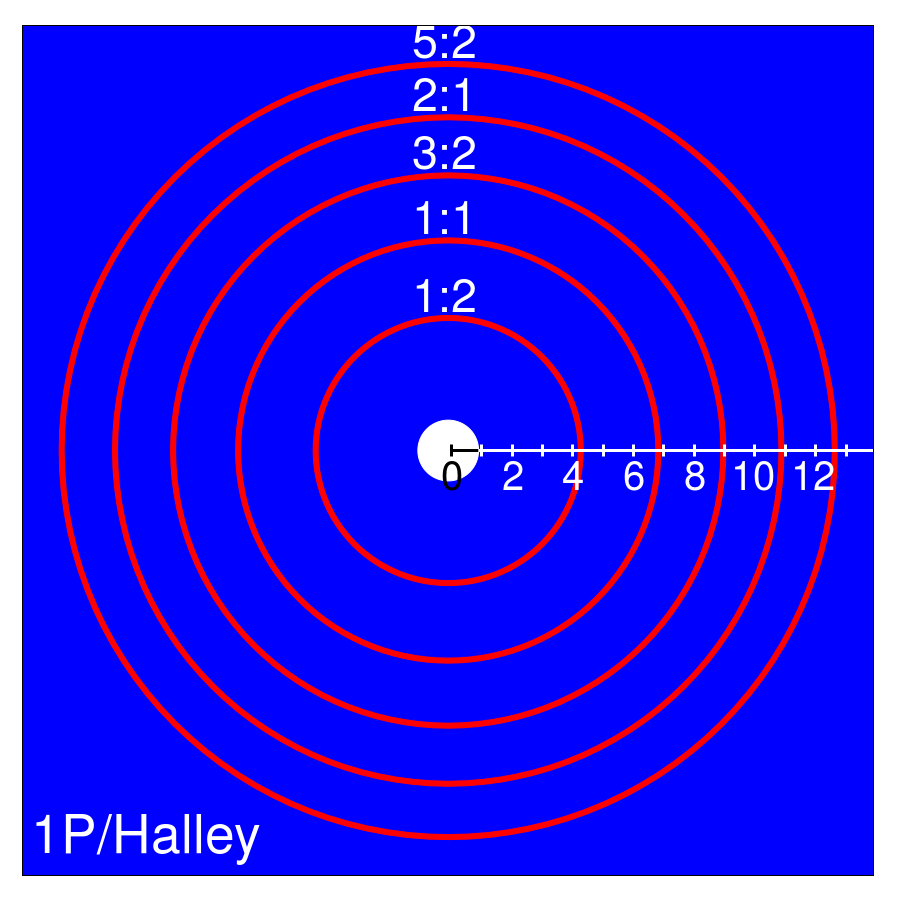}
\includegraphics[width=0.29\textwidth]{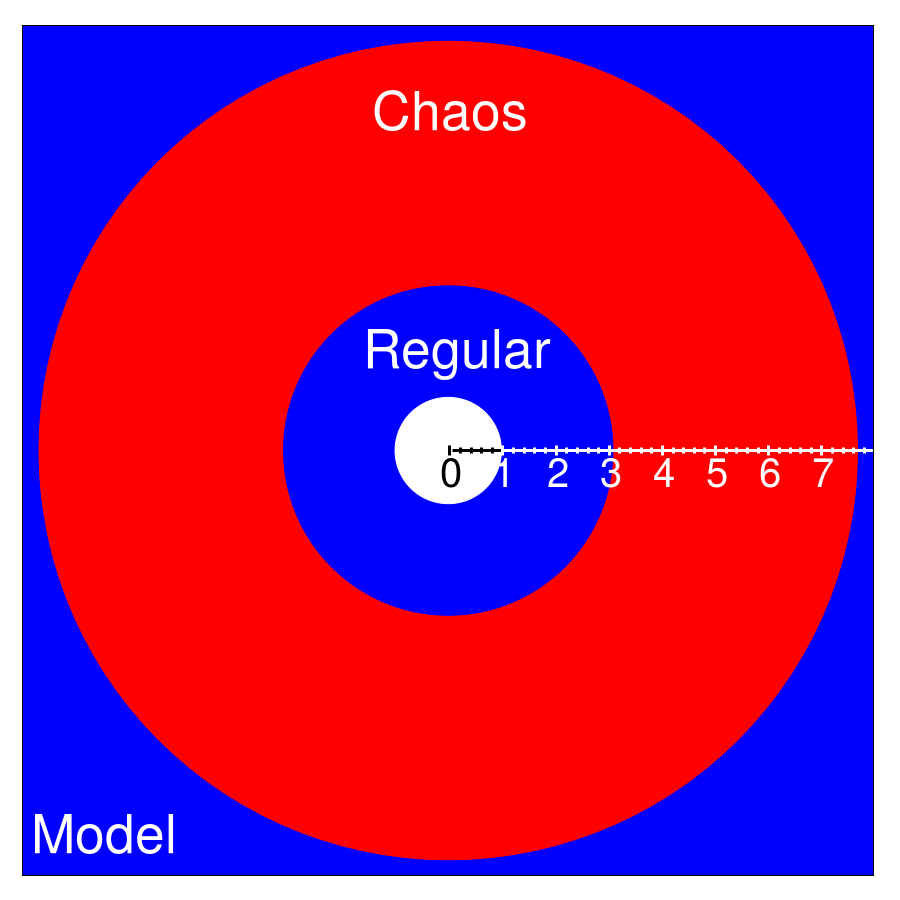}
\includegraphics[width=0.29\textwidth]{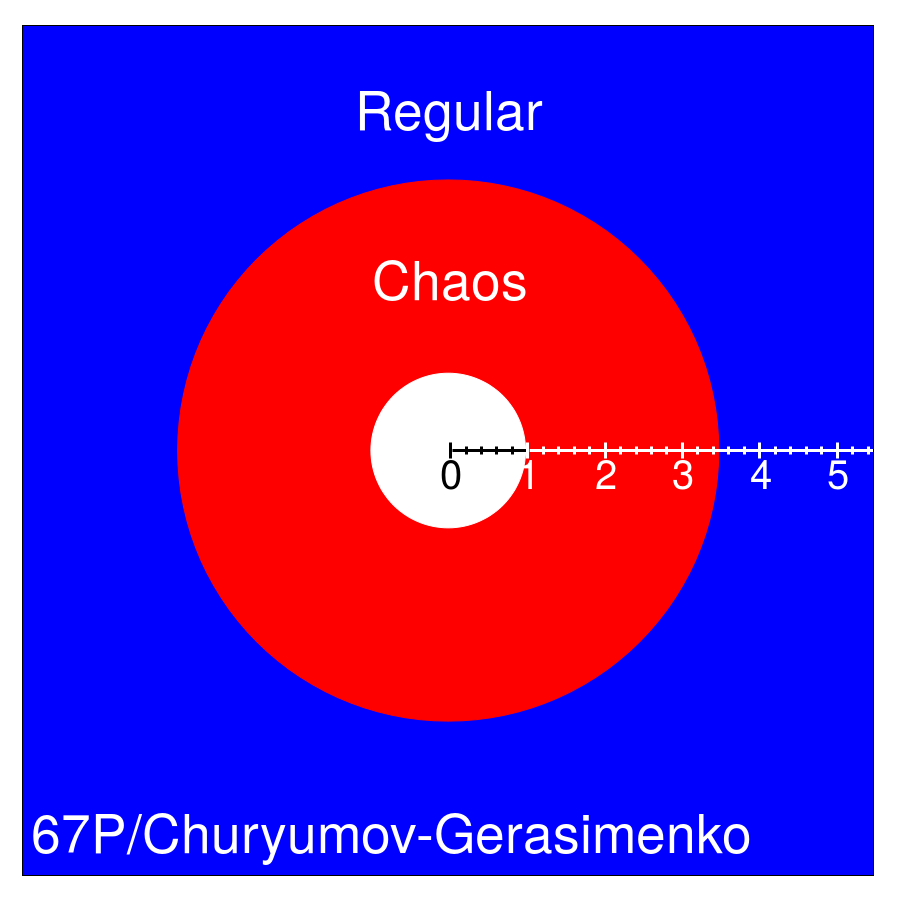}
\end{center}
\caption{Extent of the central chaotic zone around a cometary nucleus as a function of the rotation rate $\omega$.
Upper panel: the red domain corresponds to the central chaotic zone extent for the dumb-bell symmetric case $\mu=1/2$; the blue domain is the domain of stable orbits. The borders of the central chaotic zone for $\mu\simeq0.22$ (the mass parameter of \borrellyE) and $\mu\simeq0.32$ (the mass parameter of \tuttleE) are shown by the white dashed line and the white dash-dotted line, respectively. The rotation rates $\omega$ of the five cometary nuclei described in Section~\ref{dbac} are represented by vertical white lines. Locations of the integer resonance $1$:$1$ and the half-integer resonances $1$:$2$ and $2$:$1$ are represented by red lines.
Bottom panels: schematic presentations of resonances and the chaotic zones (at $e=0$) around
the nucleus of \halley (bottom left),
a model cometary nucleus with rotation rate $\omega=0.1\omega_0$ and mass parameter $\mu=0.5$ (bottom middle),
and
the nucleus of \chury (bottom right).
The centers of mass of the cometary nuclei are located at the centers of these panels.
} \label{fig2}
\end{figure}

\subsection{Circumnuclear chaotic zone}

Let us define the central chaotic zone as the region in $q$ where even the particles that start in circular orbits ($e=0$) move chaotically. 
From Fig.~\ref{fig1}, we retrieve the known fact that, $\mu$ being similar for all the cometary nuclei, the extent of the central chaotic zone significantly increases as the rotation rate $\omega$ slows down \citep{lages17}. As an illustration, here (Fig.~\ref{fig1}) the radius of the central chaotic zone ranges from $q\simeq3d$ around \hartley ($\omega\simeq1.04\omega_0$) to $q\simeq5d$ around \tuttle ($\omega\simeq0.33\omega_0$).

As the chaotic border is determined by the critical eccentricity $e_\mathrm{cr}(q)$, the solution of the equation $e_\mathrm{cr}(q)=0$ at $q>1$ can be taken as the radius of the chaotic zone around the cometary nuclei. Fig.~\ref{fig2} shows the extent of the central chaotic zone as a function of the rotation rate $\omega$.

In the $0.3\lesssim\omega/\omega_0\lesssim1$ region, which comprises the rotation rates of \tuttleE, \borrellyE, \churyE, and \hartleyE, we clearly see 
that at $\mu$ from $\approx 0.22$ to $\approx 0.32$ the extents of the chaotic zone are almost
the same
(see the white dashed and dash-dotted curves in Fig.~\ref{fig2} in the range $0.3\lesssim\omega/\omega_0\lesssim1$) and 
are even very close to the extent of the chaotic zone in the $\mu=1/2$ symmetric case. For these 4 cometary nuclei a generic illustration of the central chaotic zone is given in the bottom right panel of Fig.~\ref{fig2}.

In the $0.1\lesssim\omega/\omega_0\lesssim0.3$ region, a zone of regular orbits exists surrounding the close vicinity of the cometary nucleus which is consequently insulated from an annular chaotic zone. This configuration is explained by the fact that the $e_\mathrm{cr}(q)$ function has two roots in the $0.1\lesssim\omega/\omega_0\lesssim0.3$ region, these two roots giving the inner and outer radii of the central chaotic zone as illustrated in the bottom middle panel of Fig.~\ref{fig2} for a model cometary nucleus rotating with $\omega=0.1\omega_0$ and having the mass parameter $\mu=1/2$.

\begin{figure}
\begin{center}
\includegraphics[width=0.48\textwidth]{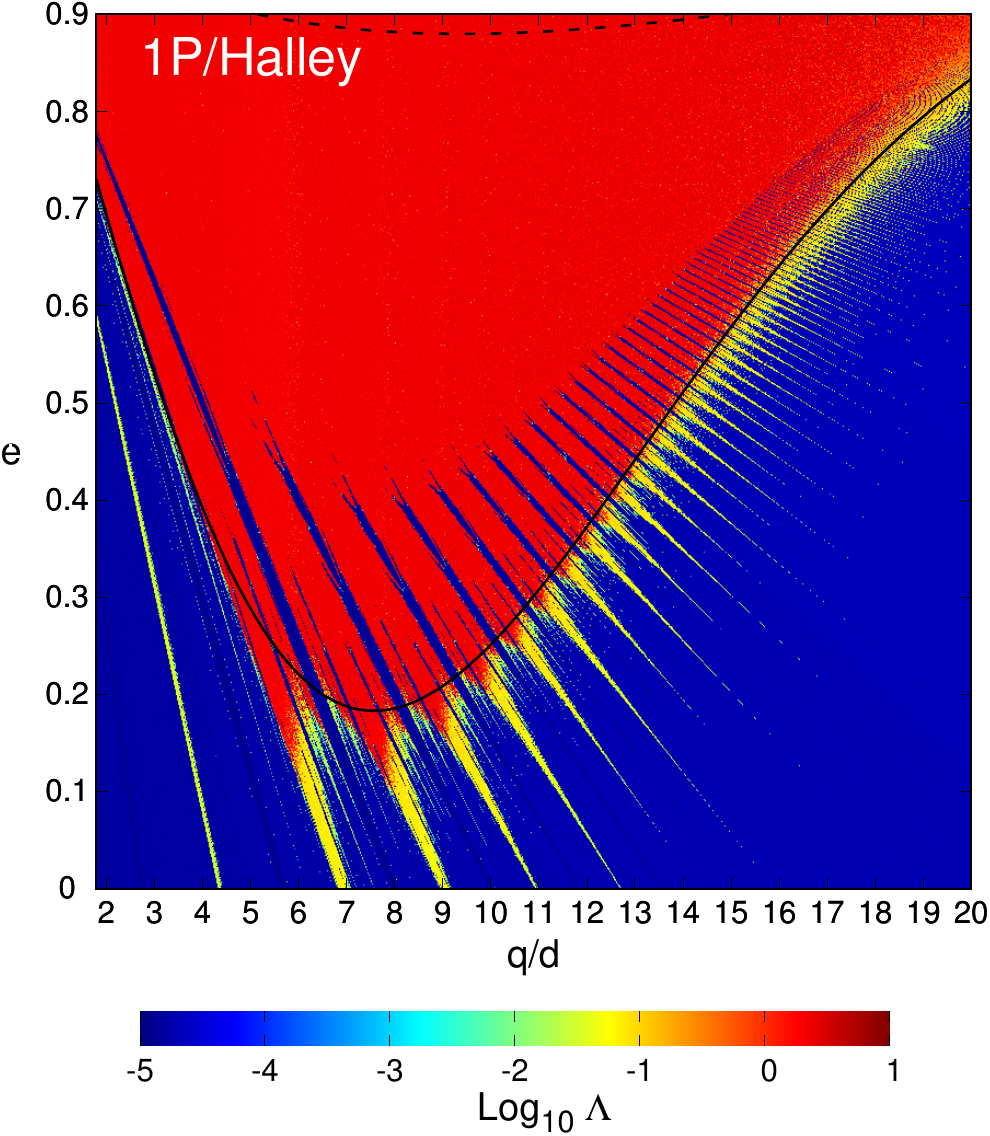}
\includegraphics[width=0.48\textwidth]{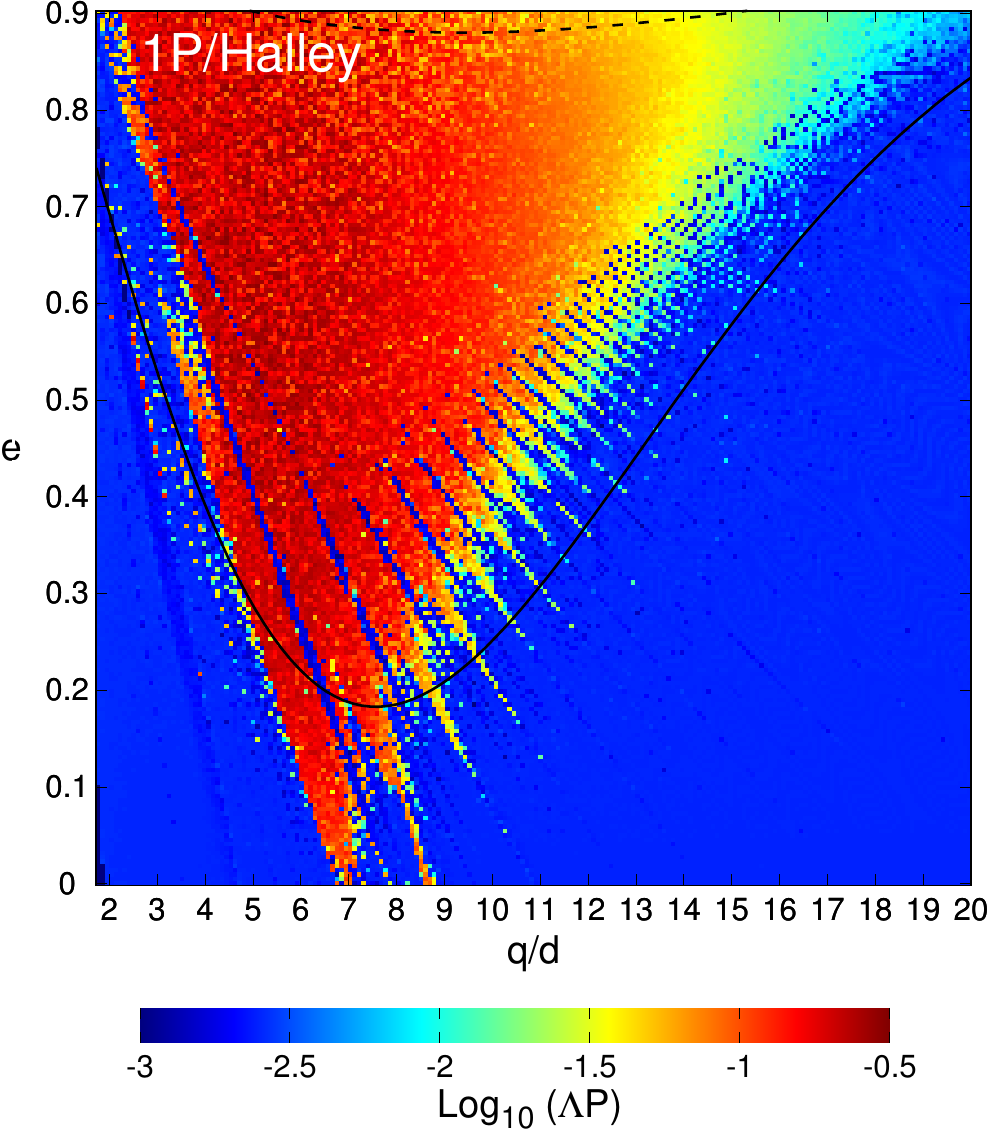}
\end{center}
\caption{Stability diagrams of the orbital motion around
\halley ($\omega\simeq0.055$, $\mu\simeq0.28$) obtained
(left panel) by iterating the dumb-bell map (\ref{kmp_gm2})
and
(right panel) by integrating the equations of motion of a particle around a contact-binary with 1P/Halley's physical characteristics.
The chaotic domain is shown by
the reddish area. Chaos is determined by computing the maximum Lyapunov
exponent $\Lambda$ for a trajectory with initial orbital elements
$(q,e)$. 
The critical curve (\ref{Km_ecr}) is shown
taking into account only the first harmonic term in $\Delta E$ (\ref{deltaEall}) with amplitude
$W_1$ (\ref{W1}) (dashed black line) and taking into account
only the second harmonic term in $\Delta E$ (\ref{deltaEall}) with amplitude $W_2$ (\ref{W2})
(solid black line).
Left panel: the number of iterations of the dumb-bell map (\ref{kmp_gm2}) is $10^6$.
The Lyapunov exponent $\Lambda$ has the physical dimension of the inverse of the number of iterations of the map.
Right panel:
the numerical integration of the equations of motion have been performed over the time duration $10^4T_0$ where $T_0=2\pi/\omega_0$.
The Lyapunov exponent $\Lambda$ has the physical dimension of the inverse of time, and $(\Lambda P)^{-1}$ expresses the Lyapunov time in the units of the rotation period $P$ of the nucleus.
} \label{fig3}
\end{figure}

In the $\omega\lesssim0.1\omega_0$ region, according to the upper panel of Fig.~\ref{fig2}, the above defined central chaotic zone does not exist. This is in accordance with the stability diagram (see Fig.~\ref{fig3}, left panel) for the particles orbiting a dumb-bell with the characteristics of the \halley nucleus ($\omega\simeq0.055\omega_0$, $\mu\simeq0.28$). In Fig.~\ref{fig3}, left panel, a particle put initially in a circular orbit has regular dynamics if the initial radius does not correspond to integer or half-integer resonances (corresponding to the resonant teeth reaching the line $e_\mathrm{cr}=0$ in Fig.~\ref{fig3}, left panel). The central chaotic zone, defined as the domain in $q$ for which any initial eccentricity implies chaos, is absent as illustrated for the \halley nucleus in the bottom left panel of Fig.~\ref{fig2}. However for higher eccentricities the chaotic domain is quite extended with e.g., a width of $\sim10d$ for $e\simeq0.5$ and $\sim20d$ for $e\simeq0.8$.

Usually the Kepler map is used to describe the highly eccentric orbital motion around a binary composed of a primary and a secondary acting as a perturber, e.g., the chaotic dynamics of \halley around the Sun and Jupiter \citep{chirikov89,rollin15b}. Also the Kepler map was derived and used for $\omega\, a^{3/2}>\omega_0\,d^{3/2}$, where $a=-1/2E$, to describe the  molecular Rydberg states with a rotating dipole core \citep{casati88,benvenuto94}. In these two cases of application, the orbital period of the tertiary is greater than the binary rotation period, and the usually considered dynamics within the Kepler map framework lies above the $1$:$1$ resonance line in Fig.~\ref{fig2}. According to Fig.~\ref{fig2} the motion of a body orbiting close to the \halley nucleus may take place below this resonance line. Consequently in order to check the relevance of application of the dumb-bell map (\ref{kmp_gm2}) below the $1$:$1$ resonance line, i.e., for $\omega\, a^{3/2}<\omega_0\,d^{3/2}$, we have numerically integrated the equations of motion of a particle around a dumb-bell with the \halley nucleus characteristics. We have obtained the corresponding stability diagram (Fig.~\ref{fig3}, right panel) which exhibits a ragged chaotic border qualitatively similar to that in the stability diagram (Fig.~\ref{fig3}, left panel) obtained iterating the dumb-bell map (\ref{kmp_gm2}). Note that in these two stability diagrams, the Lyapunov exponents are measured in different units.

\subsection{The Lyapunov times, the chaotic zone radii, and the Hill radii}
\label{czrlt}

\begin{figure}
\begin{center}
\includegraphics[width=0.48\textwidth]{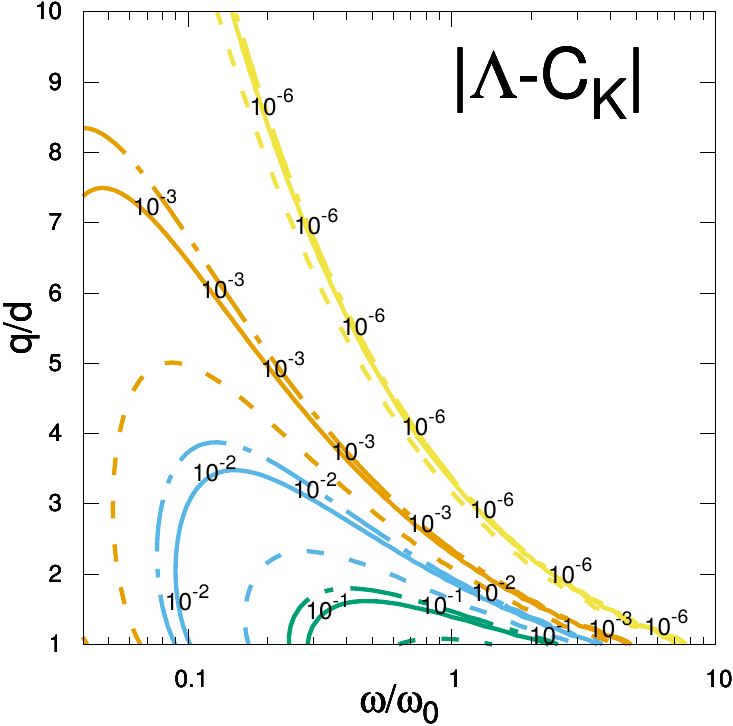}
\includegraphics[width=0.48\textwidth]{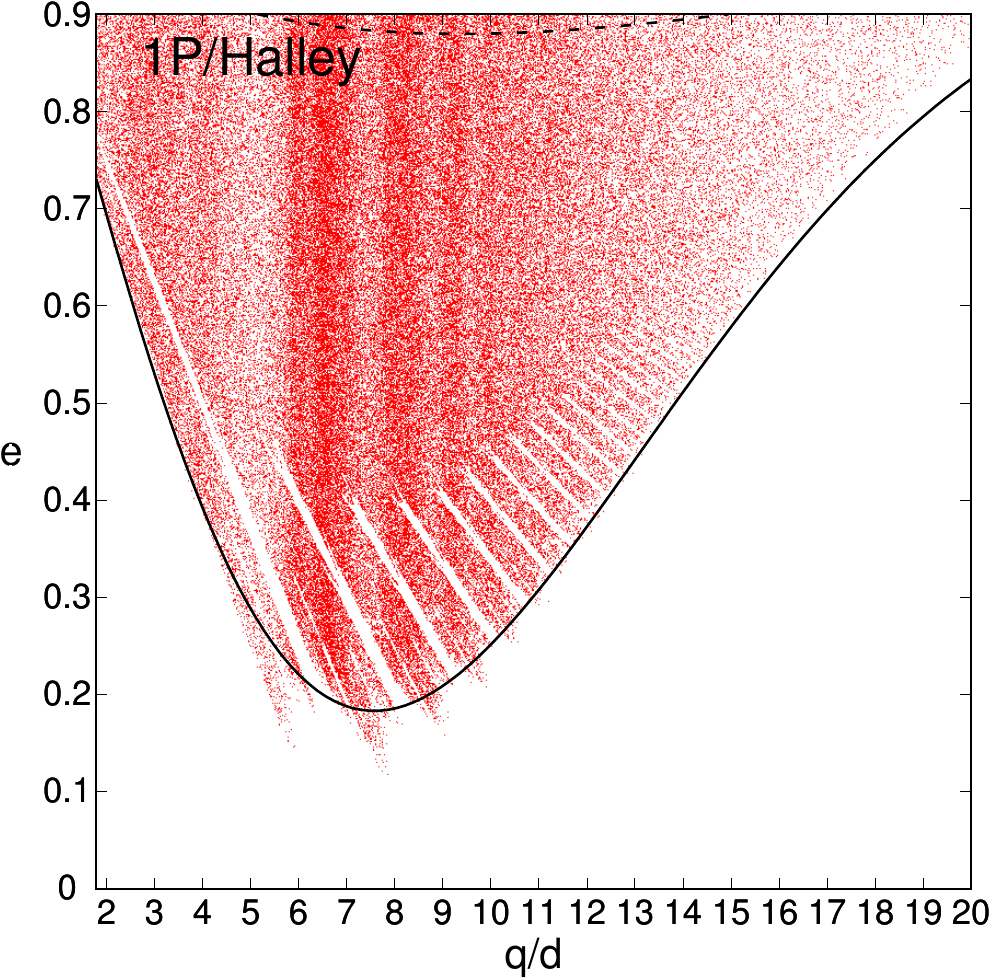}
\end{center}
\caption{Left panel: Contour plot showing the difference, $|\Lambda-C_\mathrm{K}|$, between the analytical expression of the Lyapunov exponent $\Lambda(\mu,q,\omega)$ (\ref{lambd})  and the Chirikov constant $C_\mathrm{K}\simeq2.2$; for mass parameters $\mu=0.1$ (dashed lines), $\mu=0.28$ (mass parameter of \halleyE, solid lines), and $\mu=0.5$ (dot-dashed lines). Each color is associated to a value of $|\Lambda-C_\mathrm{K}|$. Right panel: We show data from the  ``\halleyE'' stability diagram (left panel of Fig.~\ref{fig3}) but selecting only initial conditions $(e,q)$ leading to a Lyapunov exponent $2.1<\Lambda<2.3$.} \label{fig4}
\end{figure}

Roughly speaking, the inverse of the Lyapunov exponent obtained at initial conditions $(q, e)$ by iterating the dumb-bell map (\ref{kmp_gm2}) gives the number of iterations needed to observe the onset of chaos. As shown in Section~\ref{chaosborder}, the dumb-bell map (\ref{kmp_gm2}) can be rewritten as the original Kepler map (\ref{kmp_gm1}) for which the Lyapunov exponent inside the around--the--separatrix chaotic component is known \citep{shevchenko07} to be equal to
\begin{equation}
\label{lambd}
\Lambda(\lambda)=C_\mathrm{K}-3/\lambda,
\end{equation}
where $C_\mathrm{K}\simeq2.2$ is the Chirikov constant and $\lambda$ is the adiabaticity parameter of the Kepler map (\ref{kmp_gm1}). Here, in the case of rotating small bodies the adiabaticity parameter reads $\lambda=2^{1/2}\pi\omega|W_2|^{-3/2}$ \citep{lages17}.
Even in the case of \halleyE, with the slowest rotation rate considered in this study, chaos is non-adiabatic since the Kepler map parameter is still very high, $\lambda\sim\omega W_2^{-3/2}\sim100\gg1$ (from \cite{lages17} $W_2\sim10^{-2}(d\omega_0)^2$ for $\omega\simeq0.05\omega_0$ and $q>d$).
Consequently the Lyapunov exponent in the chaotic domain is expected to be approximately $\Lambda\simeq C_\mathrm{K}\simeq2.2$ for any set of the physical parameters of this study as is shown by Fig.~\ref{fig4}, left panel.
We have verified, from Fig.~\ref{fig1} and Fig.~\ref{fig3}, left panel, that the Lyapunov exponent is indeed approximately $2.2$ for the chaotic orbits around each of the five cometary nuclei (see Fig.~\ref{fig4}, right panel as an illustration for the \halley case).

The inverse of the Lyapunov exponent obtained by the direct numerical integration of Newton's equations gives, by the order of magnitude, the time (in constant natural time units) needed at given initial conditions $(e,q)$ for chaos to develop. The value of the Lyapunov exponent for an orbit with initial $(e,q)$ in Fig.~\ref{fig3} (right panel) can be straightforwardly estimated using the corresponding value from Fig.~\ref{fig3} (left panel) divided by the corresponding mean orbital period.

\begin{table}
\begin{center}
\caption{\label{Table_res}
Bilobed cometary nuclei: sizes of chaotic zones and the Lyapunov
times}
\begin{threeparttable}
\begin{tabular}{lrrrrrrrr}
\noalign{\medskip} \hline \noalign{\medskip}
Comet & $\omega/\omega_0$ & \multicolumn{2}{c}{$R_\mathrm{ch}$, km} & $R_\mathrm{Hill}$, km & \multicolumn{2}{c}{$R_\mathrm{ch}/R_\mathrm{Hill}$, \%}&$T_\mathrm{L}$, h &$T_\mathrm{L}/P$\\
&&$e\simeq0$&$e\simeq0.5$&&$e\simeq0$&$e\simeq0.5$&&\\
\noalign{\medskip} \hline\hline \noalign{\medskip}
67P/C--G	& 0.73	& 9	& 11			& 320	& 3\%	& 3.3\%		& 35	& 2.9	\\
\halley		& 0.055	& --	& 31-108\tnote{ a}	& 200	& --	& 16-54\%	& 230	& 1.3	\\
\tuttle		& 0.33	& 25	& 32.5 			& 620	& 4\%	& 5.2\% 	& 28	& 2.5	\\
\borrelly	& 0.48	& 16	& 21 			& 300	& 5\%	& 7\%		& 66	& 2.6	\\
\hartley   	& 1.04	& 3.4	& 4.2			& 52	& 7\%	& 8.1\%		& 56	& 3.1	\\
\noalign{\smallskip} \hline
\end{tabular}
\begin{tablenotes}
\item[a]{\small For the \halley nucleus we give the inner and outer radii of the annular chaotic zone at $e\simeq0.5$.}
\end{tablenotes}
\end{threeparttable}
\end{center}
\end{table}

Table~\ref{Table_res} summarizes the radii of the chaotic zone $R_\mathrm{ch}$ assessed directly from the $(e,q)$ stability diagrams for the cometary nuclei (Figs.~\ref{fig1} and \ref{fig3}).
Except the case of \halleyE, the chaotic zone radii at $e\simeq0$ and at $e\simeq0.5$ are similar in each case.

Let us compare these radii to the corresponding Hill radii. In the planar circular restricted three-body problem, the Hill
radius is given by $R_\mathrm{Hill} \approx a_\odot\left( M/ 3M_\odot \right)^{1/3}$
where $a_\odot$ is the semi-major axis of a comet's orbit around the Sun. In the elliptic
problem, a ``pericenter scaling'' for $R_\mathrm{Hill}$ is given
by $R_\mathrm{Hill} \approx q_\odot\left( M/3 M_\odot\right)^{1/3}$, where $q_\odot$ is a comet's perihelion distance \citep{hamilton92}. 
The calculated Hill radii $R_\mathrm{Hill}$ are presented in Table~\ref{Table_res}.
From Table~\ref{Table_res}, we see that that the typical size of \halleyE's
chaotic zone $R_\mathrm{ch}$ is the largest one in the sample, and
it seems to engulf an essential part of the Hill sphere, at least
for the orbits of moderate and high eccentricity (38\% of the chaotic zone overlaps with the Hill sphere at $e\simeq0.5$). This is an outcome of the
slowness of rotation of the nucleus.

As mentioned above, $\omega \gtrsim \omega_0$ are not generally probable, as a
rubble-pile object would disintegrate at such high rates of
rotation.
For Comet 103P/Hartley, $\omega \approx \omega_0$; therefore, the nucleus may formally be on the brink of disintegration. The chaotic zone of this object is formally minimal in size in the sense that at $\omega < \omega_0$ the zone would be larger.

Following the general approach presented in
\cite[section~5]{shevchenko07}, the local Lyapunov time
$T_\mathrm{L}$ can be estimated, for a satellite assumed to move
within the chaotic domain around a nucleus. It is approximately
equal to the ratio of the satellite's orbital period and the map's
Lyapunov exponent. The map's Lyapunov exponent averaged over the chaotic layer does not depend
practically on the adiabaticity parameter, if the latter is high enough; see \cite{shevchenko07}.
The local Lyapunov time is estimated taking the satellite's
orbital period $T_\mathrm{orb}$ corresponding to the radius of the central chaotic zone $(a = R_\mathrm{ch})$ at $e \simeq 0.5$ as given in Table~\ref{Table_res}, as this gives the maximum possible $T_\mathrm{orb}$ for the chaotic motion at zero and moderate eccentricities, and, consequently, the maximum possible local Lyapunov time. These estimations of the typical local Lyapunov time for each cometary nucleus are given in Table~\ref{Table_res}.
One can readily see that the  Lyapunov times do not differ substantially, ranging from $\sim 1$~d to $\sim 10$~d. The relative range is even less, if the  Lyapunov times are measured in the units of the rotation period of the cometary nuclei; then they range from $1.3$ to $3.1$.
For the satellites with smaller orbits ($a<R_\mathrm{ch}$), the local Lyapunov time is smaller, and, consequently, the motion is less predictable.

Among the five objects listed in Table~\ref{Table_data}, four out of the five had been visited and observed closely in their vicinities by space probes:  
1P/Halley (by Vega-1, Vega-2, Giotto in 1986), 
19P/Borrelly (by Deep Space 1 in 2001), 
67P/C--G (by Rosetta 1 in 2014--16), and 
103P/Hartley (by Deep Impact EPOXI mission in 2010). 
No satellites of the nuclei had been observed (apart from apparently replenishable clouds of  ``grains'') during these close rendezvous, although meticulous surveys had been done in some cases; see \cite{bertini15} and references therein. The observed absence of cometary satellites is apparently in agreement with our theoretical findings on the large extents of the circumnuclear chaotic zones. 

Among these four nuclei, two objects were observed in such a detail as to allow for considering the close-to-nucleus mass transfer and dynamics of various ejecta, which include, in particular, quite large (decimetre sized) chunks of water ice \citep{ahearn11,keller14}. Thus, there exists a source of material which can be introduced in orbits around nuclei, as theoretically described in \cite{fulle97} and \cite{scheeres00}. However, our theoretical findings imply that, even in the absence of any forces other than gravitational, no long-lived orbits of any material may sustain inside the circumnuclear chaotic zones.  

One may say that the rotating irregularly-shaped nuclei clean up their vicinities. This clearing is analogous in some way to the clearing of the ``Wisdom gap'' in the coorbital vicinities of a planet that is massive enough. (On Wisdom's coorbital chaotic layer, see, e.g., \citealt{murray99}.) The dynamical difference is that the Wisdom gap is formed by the overlap of accumulating first-order mean motion resonances in the coorbital vicinity of a planet, whereas the circumnuclear chaotic zone is formed by the overlap of accumulating integer and half-integer orbit-spin resonances with the rotating nucleus, as discussed in Section~\ref{stabdiag}. 

Therefore, our theoretical expectation is that no long-lived satellites, or any long-lived non-replenishable halos formed of large particles, can be normally observed to exist around bilobed cometary nuclei, within the defined boundaries of the circumnuclear chaotic zone. 

\section{Conclusions}

We have used the generalized ``dumb-bell'' Kepler map,
introduced recently in \cite{lages17}, to describe the
qualitative chaotic dynamics around bilobed cometary nuclei. The
analysis has been based on the data for five comets whose nuclei
are well-documented to resemble dumb-bells. As lobes' masses are comparable, $m_1\sim m_2$, the chaotic zone's configuration and extents depend mostly on the cometary nuclei rotation rate $\omega$.
The sizes of chaotic
zones around the nuclei and the Lyapunov times of the motion
inside these zones have been estimated. In the case of Comet
\halleyE, the chaotic zone seems to engulf an essential part of
the Hill sphere, at least for orbits of moderate to high eccentricity.

Therefore, simple analytical formulas, based on the Kepler map formalism, allow one to straightforwardly estimate the sizes of chaotic zones around the dumb-bell cometary nuclei and the Lyapunov times of the motion inside these zones.

In practice, the obtained numerical estimates of the characteristics of the chaotic zones around the five considered objects allow one to judge where (in the space of orbital parameters) any small natural satellites cannot be expected to orbit any of these objects, or where any artificial satellite cannot be put in a stable orbit. On the other hand, the knowledge of the Lyapunov times allows one to judge on which timescales the coordinates of a satellite orbiting in the chaotic zone are predictable.

\section*{Acknowledgements}

We are most thankful to the referees for important remarks.
We are grateful to Dima L. Shepelyansky for valuable remarks, comments, and discussions. I.I.S. benefited from a grant of
Bourgogne-Franche-Comt\'e region. I.I.S. was supported in part by
the Russian Foundation for Basic Research (project No.
17-02-00028)
and the Programmes of Fundamental Research of the Russian Academy
of Sciences ``Fundamental Problems of Nonlinear Dynamics'' and
``Fundamental Problems of the Solar System Study and
Exploration''.

\section*{References}

\bibliography{CDCNbib}{}

\end{document}